\documentclass[dvips,12pt]{article}
\usepackage{graphicx,color}
\newcommand\figcaption{\def\@captype{figure}\caption}
\newcommand\tabcaption{\def\@captype{table}\caption}
\def\pa{\partial}
\def\al{\alpha}
\def\ga{\gamma}
\def\Ga{\Gamma}
\def\dl{\delta}
\def\Dl{\Delta}
\def\la{\lambda}

\def\be{\beta}
\def\kp{\kappa}

\def\sg{\sigma}
\def\sg{\sigma}

\def\nb{\nabla}

\def\nn{\nonumber}
\def\ck{\check}
\def\wt{\widetilde}
\def\diag{\mbox {diag}}
\title{\LARGE\bf Local Lorentz Transformation and \\``Lorentz Violation"}
\author{Ying-Qiu Gu\footnote {email: yqgu@fudan.edu.cn}}
\date{\small School of Mathematical Science, Fudan University, Shanghai 200433, China}

\begin{document}
\maketitle\DeclareGraphicsRule{.eps.gz}{eps}{.eps.bb}{`gunzip -c
#1}

\begin{abstract}
Some solutions to the anomalies of ultra high energy
cosmic-ray(UHECR) and TeV $\gamma$-rays require disturbed
non-quadratic dispersion relations, which suggest the Lorentz
violation. Also, some new theories such as quantum gravity, string
theory and the standard model extension imply the Lorentz
violation. In this paper, we derive some transformation laws for
the classical parameters of nonlinear field system, and then
provide their dispersion relations. These dispersion relations
also have non-quadratic form but keep the Lorentz invariance. The
analysis of this paper may be helpful for understanding the
quantum theory and the plausible Lorentz violation.

\end{abstract}

\section{Introduction}
\setcounter{equation}{0}

Recently the ``Lorentz violation" becomes a hot topic. This
situation arises from both experimental and theoretical
developments. One is the astronomical observation of the anomalies
of ultra high energy cosmic-ray(UHECR) and the TeV $\ga$-rays. The
other is the implication of some new physics such as the extended
gauge theory, quantum gravity and string theory. In observational
aspect, the UHECR protons with energy $>10^{20}$eV were
observed\cite{e1,e2,e3}. The energy of these protons are beyond
the GZK cutoff $5\times 10^{19}$eV, which was derived
independently by Greisen\cite{e4} and Zatsepin , Kuz¡¯min\cite{e5}
shortly after the discovery of the cosmic microwave background
radiation(CMBR). According to the particle theory, the cosmic-ray
nucleons with sufficient energy will inelastic collide with
photons of CMBR to produce baryons or pions as follows
\begin{equation}
P+\ga({\mbox{CMB}})\to \Dl,\qquad P+\ga({\mbox{CMB}})\to p+ N \pi,
\label{1.1} \end{equation} so the nucleons with energies beyond
the GZK limit cannot reach us from a source further than a few
dozen Mpc. It should be mentioned that, this GZK limit and the
physical process are well understood and measured in the
laboratory\cite{e6}.

Another similar paradox is the TeV $\ga$-rays.  Two photons with
energy over $2m_e c^2$ can produce an electron-positron pair
$\ga+\ga \to e^-+e^+$. Photons of 10 TeV are most sensitive to $30
\mu m$ infrared photons, so a photon with enough energy
propagating in the intergalactic space will interact with infrared
background photons, and be exponentially suppressed. However the
10 TeV photons from Mkn501, a BL Lac objects at distance about
150Mpc\cite{e7,t1}, could reach the Earth.

To explain the above anomalies, numerous solutions have been
proposed. Most of these solutions are related to slight violation
of Lorentz invariance to get a shift of the thresholds of the
energy. Glashow believe that the limit speed of a particle depends
on its species and should be the eigenvalue of the velocity
eigenstates\cite{gls1}-\cite{gls7}. Then for particle $P$ the
dispersion relation becomes
\begin{equation}
E^2=p^2c_P^2+m_P^2c_P^4. \label{1.3} \end{equation} In
\cite{ncg1}, the authors suggested a non-quadratic dispersion
relation for photons
\begin{equation}
p^2 c^2 = E^2 (1 + f\left(\dl)\right),\quad \dl=\frac E
{E_{QG}}\to 0\label{1.4} \end{equation} where $E_{QG}\sim
10^{19}$GeV is an energy scale caused by quantum gravity effect.
Then the light speed is perturbed by E as
\begin{equation}\wt c=\frac {\pa E}{\pa p}\approx c(1+k\dl).\label{1.5} \end{equation}
This assumption causes the Lorentz violation and deformed special
relativity, which is related to the $\kp$-Poincar\'e
superalgebra\cite{ncg2}. The Casimir of the $\kp$-Poincar\'e
superalgebra has a structure similar to (\ref{1.4}). Nowadays, the
deformed relativity or noncommutative geometry is greatly
developed\cite{ncg3}-\cite{ncg11}.

In theoretical aspect\cite{LV,sme,sme1}, the Standard Model
Extension and quantum gravity suggest that Lorentz invariance may
not be an exact symmetry. The possibility Lorentz violation has
been investigated in different quantum gravity models, including
string theory\cite{sme2,sme3}, warped brane worlds\cite{sme4}, and
loop quantum gravity\cite{sme5}. These models adopt the Lagrangian
like the following\cite{sme6}
\begin{equation}
{\cal L}=\bar \psi\left(\frac 1 2 e^\mu_{~a}\Ga^a \stackrel
{\leftrightarrow}{\pa}_\mu-M\right)\psi, \label{1.6}
\end{equation} where $\psi$ is Dirac spinor, $e^{\mu}_{~a}$ is the
vierbein, and $e^{~a}_{\mu}$ is the inverse,
\begin{eqnarray}
\Ga^a&=&\ga^a-c_{\mu\nu}e^{\nu a}e^\mu_{~b}\ga^b-f_\mu e^{\mu a}-i
k_\mu e^{\mu a}\ga_5+\cdots, \label{1.7}\\
M&=& m+i\mu \ga_5 +a_\mu e^\mu_{~a}\ga^a+b_\mu
e^\mu_{~a}\ga^a\ga_5+\cdots. \label{1.8} \end{eqnarray} The first
right terms of (\ref{1.7}) and (\ref{1.8}) correspond to the
normal Lorentz invariant kinetic term and mass for the Dirac
spinor. But the other coefficients are Lorentz violating
coefficients arising from nonzero vacuum expectation values of the
coupling tensor fields, which seem to be introduced quite
arbitrarily.

In contrast with the above theories of Lorentz violation, the
torsion theory is the most natural one in logic\cite{tsn1,tsn2},
which is derived from the fact that the connection can compatibly
introduce an antisymmetric part, namely, the torsion.
\begin{equation}
\wt \Ga^\mu_{\al\be}=\Ga^\mu_{\al\be}+T^\mu_{\al\be}, \label{1.9}
\end{equation} where $\Ga^\mu_{\al\be}$ is the Christoffel symbol,
and $T^\mu_{\al\be}=-T^\mu_{\be\al}$ is the torsion of the
spacetime. Different from all matter fields like electromagnetic
field, torsion is a geometrical interaction similar to gravity,
which uniformly interacts with all matter in an accumulating
manner. So to test torsion, it seems more effective to measure the
movement of heaven body rather than atoms.

However, in contrast with the natural essence and deep
philosophical meanings of the Lorentz invariance, the violation
theories seem to be somehow artificial\cite{obj}. Some experiments
have been elaborated to test the Lorentz violation, but all
results gave negative answers in high
accuracy\cite{tst1}-\cite{tst7}, so one should take a little
conservative attitude towards the Lorentz violation. Then how to
explain the threshold anomalies of UHECR and TeV $\ga$-rays? Here
we present another scenario based on the nonlinear field theory,
which also provides non-quadratic mass-energy relations or
dispersion relations similar to (\ref{1.4}), but strictly keeps
the Lorentz invariance\cite{gu1,gu2,gu3}. In what follows we
examine the local Lorentz transformation for some classical
parameters defined from nonlinear fields and establish their
relations.

\section{Local Lorentz transformation and non-quadratic dispersion relation}
\setcounter{equation}{0}

\subsection{Local Lorentz transformation for classical parameters}

Taking the Minkowski metric as $\eta_{\mu\nu}={\rm
diag}[1,-1,-1,-1]$, we consider the field systems of nonlinear
spinor $\psi$ and scalar $\phi$. For the nonlinear spinor $\psi$,
the dynamic equation is given by\cite{sln1}-\cite{gu6}
\begin{equation}
\al^\mu (\hbar i\pa_\mu-eA_\mu)\psi=(\mu-F')\ga\psi,
\label{2.1}\end{equation} where the $4\times4$ Hermitian matrices
are defined by
\begin{equation}\al^\mu=\left\{\left ( \begin{array}{ll} I & ~0 \\
0 & I \end{array} \right),\left (\begin{array}{ll} 0 & \vec\sg \\
\vec\sg & 0 \end{array}
\right)\right\},~ \ga =\left ( \begin{array}{ll} I & ~0 \\
0 & -I \end{array} \right)\label{2.2}
\end{equation}
with Pauli matrices
\begin{equation}
 {\vec\sg}=(\sg^k)= \left \{\pmatrix{
 0 & 1 \cr 1 & 0},\pmatrix{
 0 & -i \cr i & 0},\pmatrix{
 1 & 0 \cr 0 & -1}
 \right\}.\label{2.3}\end{equation}
$F=F(\ck\ga)$ is a positive function of the quadratic scalar
$\ck\ga \equiv \psi^+\ga\psi$.

For scalar field $\phi(x^\mu)$, the dynamic equation is given by
\begin{equation}
\pa_\mu\pa^\mu \phi=K\phi,\label{2.4}
\end{equation}
where $K=K(|\phi |^2,\pa_\mu\phi^+\pa^\mu\phi)$ is a smooth real
function.

For both spinor $\psi$ and scalar $\phi$, the current conservation
law holds due to the gauge invariance of their dynamic equations,
\begin{equation}\pa_\mu\rho^\mu=0,\label{2.5}\end{equation}
where the current is defined respectively by
\begin{equation}\rho^\mu=\left\{\begin{array}{ll}
\psi^+\al^\mu\psi~~&{\mbox{for spinor}},\\
{i}\kp{{\Im}}(\phi^+\pa^\mu\phi)~~&{\mbox{for
scalar}}.\end{array}\right.\label{2.6}
\end{equation} $\kp$ is a normalizing constant. By (\ref{2.5}) we have the normalizing condition
\begin{equation}\int_{R^3}\rho^0 d^3 x=1. \label{2.7}\end{equation}

For the nonlinear equations (\ref{2.1}) and (\ref{2.4}), their
solutions have particle-wave duality\cite{sln1}-\cite{gu6}. In
\cite{gu2,gu3,gu7,gu8}, the local Lorentz transformations were
widely used for the classical parameters without proof. Here we
set the transformations on a solid base at first, and then derive
the non-quadratic dispersion relations for some cases. The
conditions for these results are helpful to understand the
relation between classical mechanics and quantum theory. To
clarify the status of the field system, we defined

{\bf Definition 1~} For the field system $\psi$ or $\phi$, we
define the central {\bf coordinate} $\vec X$ and drifting {\bf
speed} $\vec v$ respectively by
\begin{equation}\vec X(t)=\int_{R^3} \vec x \rho^0 d^3 x,\qquad \vec v=\frac d {dt} \vec
X,\label{2.8} \end{equation} where $t=x^0$. The coordinate system
with central coordinate $\vec X=0$ is called {\bf the central
coordinate system} of the field.

{\bf Definition 2~} If a field is a localized wave pack drifting
smoothly without emitting and absorbing energy quantum, that is,
it is at the energy eigenstate in its central coordinate system,
we call it at the {\bf particle state}. Otherwise, the field is in
the process of exchanging energy quantum with its environment, we
call it in the {\bf quantum process}.

By the current conservation law (\ref{2.5}), we have
\begin{equation}
\vec v=\int_{R^3} \vec x\pa_0\rho^0d^3 x=-\int_{R^3} \vec x
\nb\cdot \vec \rho d^3x =\int_{R^3} \vec \rho d^3x.\label{2.9}
\end{equation}
For the field at particle state with mean radius much less than
the characteristic length of its environment, by (\ref{2.8}) and
(\ref{2.9}) we have the classical approximation
\begin{equation} \rho^\mu\to u^\mu
\sqrt{1-v^2}\dl(\vec x-\vec X), \label{2.10} \end{equation} where
\begin{equation}u^\mu\equiv (\xi , \xi \vec v),
\quad \xi=\frac 1 {\sqrt{1-v^2}}.\label{2.11} \end{equation}
(\ref{2.10}) is the precondition for validity of classical
mechanics\cite{gu2,gu3,gu7}. For such system at particle state, we
can clearly define the classical parameters such as ``momentum",
``energy" and ``mass", and derive the classical mechanics. From
\cite{gu2,gu3}, we learn that a system at particle state can be
described by classical mechanics in high accuracy, whereas for the
system in the quantum process, we must describe it by quantum
theory or by the original equation (\ref{2.1}) or (\ref{2.4}). The
quantum process is an unstable state, which is usually completed
in a very short time.

In what follows, we examine the local Lorentz transformation for
the classical parameters. Since the rotation transformation is
trivial, we only consider the boost one. For the case of flat
spacetime, assume $x^\mu$ is the Cartesian coordinate. Consider
the central coordinate system of the field with coordinate $\bar
X^\mu$, which moves along $x^1$ at speed $v$ with $\bar
X^k(k\ne0)$ parallel to $x^k$, and $\bar X^k=0$ corresponds to the
central coordinate of the field $X^k(t)$. Then the Lorentz
transformation between $x^\mu$ and $\bar X^\mu$ in the form of
matrix is given by
\begin{equation}x=L(v)\bar X,\quad \bar X=L(v)^{-1} x=L(-v)
x\label{2.12} \end{equation} where $x=(t, x^1, x^2, x^3)^T$, $\bar
X=(\bar X^0, \bar X^1, \bar X^2, \bar X^3)^T$ and
\begin{eqnarray}
L(v)=\diag \left[\pmatrix{
 \xi & \xi v \cr \xi v & \xi},1,1\right]=(L^\mu_{~\nu}).\label{2.13} \end{eqnarray}

Assume $S, P^\mu$ and $T^{\mu\nu}$ are any scalar, vector and
tensor defined by the real functions of $\phi$ or $\psi$ and their
derivatives, such as $S=|\phi|^2$,
$T^{\mu\nu}=\Re(\psi^+\al^\mu\pa^\nu\psi)$. For the field at
particle state, all these functions are independent of proper time
$\bar X^0$. Thus in the central coordinate system, the spatial
integrals of these functions define the proper classical
parameters of the field, and these proper parameters are all
constants. Their Lorentz transformation laws are given by

{\bf Theorem 1~} {\em For a field system at particle state, the
integrals of covariant functions $S, P^\mu$ and $T^{\mu\nu}$
satisfy the following instantaneous Lorentz transformation laws
under the boost transformation (\ref{2.12}) between $x^\mu$ and
$\bar X^\mu$ at $d t=0$,
\begin{eqnarray}
I &\equiv& \int_{R^3} S(x) d^3x = \sqrt{1-v^2} \bar I. \label{2.14}\\
I^\mu &\equiv& \int_{R^3} P^\mu(x) d^3x = \sqrt{1-v^2}
L^\mu_{~\nu} \bar I^\nu. \label{2.15}\\
I^{\mu\nu}&\equiv& \int_{R^3} T^{\mu\nu}(x) d^3x =
\sqrt{1-v^2}L^\mu_{~\al}L^\nu_{~\be}\bar I^{\al\be}. \label{2.16}
\end{eqnarray} Where $\bar I, \bar I^\mu, \bar I^{\mu\nu}$ are the
proper parameters defined in the central system}
\begin{eqnarray}
\bar I =\int_{R^3} S(\bar X) d^3\bar X, \quad \bar I^\mu =
\int_{R^3} \bar P^\mu d^3\bar X, \quad \bar I^{\mu\nu}= \int_{R^3}
\bar T^{\mu\nu} d^3\bar X. \label{2.17} \end{eqnarray}

{\bf Proof~} We take (\ref{2.15}) as example to show the
relations. For the field at the particle state, by the
transformation law of vector, we have
\begin{equation}P^\mu (x)=L^\mu_{~\nu}\bar P^\nu (\bar X)=L^\mu_{~\nu}\bar P^\nu (\bar X^1,\bar X^2,\bar
X^3)=P^\mu (\xi(x^1-v t),x^2,x^3). \label{2.18} \end{equation} So
the integral can be calculated as follows
\begin{eqnarray}
I^\mu &=& \int_{R^3} P^\mu (x) d^3x\left|_{d t=0}\right.\nn\\
& = & \int_{R^3} P^\mu (\xi(x^1-v
t),x^2,x^3) \sqrt{1-v^2}d(\xi(x^1-vt))dx^2 dx^3\label{2.19}  \\
&=& \int_{R^3} L^\mu_{~\nu}\bar  P^\nu (\bar X) \sqrt{1-v^2} d^3
\bar X=\sqrt{1-v^2} L^\mu_{~\nu}\bar I^\nu.\nn
\end{eqnarray} The proof is finished.

{\bf Remarks 1~}{\em The Lorentz transformation laws (\ref{2.14}),
(\ref{2.15}) and (\ref{2.16}) are valid for the varying speed
$v(t)$, because the integrals are only related to the simultaneous
condition $d x^0=dt=0$, and the relations only related to
algebraic calculations.}

{\bf Remarks 2~} {\em  When the field is not at the particle
state, the covariant integrands will depend on the proper time
$\bar X^0$, then the calculation (\ref{2.19}) can not pass
through, and the relations (\ref{2.14})-(\ref{2.16}) are usually
invalid, except the integrand satisfies some conservation law
similar to (\ref{2.5}).}

In some text books, the relations (\ref{2.14})-(\ref{2.16}) are
directly derived via Lorentz transformation of the integrands and
volume element relation
\begin{equation} d^3x=\sqrt{1-v^2}d^3 \bar
X.\label{2.20} \end{equation} As mentioned by remark 2, the
calculation will provides wrong result if the field is not at the
particle state, because (\ref{2.20}) is just spatial volume, it
suffers from the problem of simultaneity.

Usually, the proper parameters have very simple form.  For any
true vector, it always takes $\bar I^\mu=(\bar I^0,0,0,0)$, then
\begin{eqnarray}
I^{\mu}=\sqrt{1-v^2}\bar I^0 u^{\mu}.\label{2.21} \end{eqnarray}
For some cases of tensor $ I^{\mu\nu}$, it can be expressed as
\begin{eqnarray}
I^{\mu\nu}=\sqrt{1-v^2}\left(\bar K u^\mu u^\nu+\bar
J\eta^{\mu\nu}\right),\label{2.22} \end{eqnarray} where $\bar K,
\bar J$ are constants.

In the curved spacetime with orthogonal time coordinate, if the
radius of curvature in the neighborhood of the center of the field
is much larger than the mean radius of the field, the above
calculations and relations can be parallel transformed into the
curved spacetime. In this case, let $\bar X^\mu$ be the central
Cartesian coordinate of the tangent spacetime at the center of the
field. $X^\mu$ is an inertial coordinate system in the tangent
spacetime fixed with the curved spacetime. $\bar X^\mu$
instantaneously moves along $X^1$ at speed $v$. Then it is easy to
check the local Lorentz transformation laws (\ref{2.14}),
(\ref{2.15}) and (\ref{2.16}) also hold, as long as the field is
at the particle state.

\subsection{Non-quadratic dispersion relation for spinor}

For the dark spinor,  $e=0$ in (\ref{2.1}). From \cite{gu2,gu3},
we get the momentum and mass-energy relation of the spinor as
follows
\begin{eqnarray}
p^{\mu}&=&\left(m_0+W{\rm ln}\frac 1 {\sqrt{1-v^2}}\right)u^{\mu}, \\
E&=&\frac{m_0}{\sqrt{1-v^2}}+W\left(\frac 1 {\sqrt{1-v^2}} {\rm
ln} \frac 1 {\sqrt{1-v^2}}+{\sqrt{1-v^2}}\right),
 \label{2.23} \end{eqnarray}
where $W\ll m_0$ is the proper energy provided by the nonlinear
potential. The mass-energy relation or dispersion relation in the
usual form is given by
\begin{eqnarray}
\left(E-W\sqrt{1-v^2}\right)^2=\vec p^{~2}+\left(m_0+W{\rm
ln}\frac 1{\sqrt{1-v^2}}\right)^2.\label{2.24}\end{eqnarray}
Referring to electron, by estimation we have $\frac W {m_0}\sim
10^{-6}$. For a spinor moving at speed $1-10^{-n}$, for $n>1$ we
have
\begin{equation}
{\rm ln} \frac 1 {\sqrt{1-v^2}}\approx
1.15n-0.35.\label{2.26}\end{equation} The Lagrangian of the
particle becomes
\begin{equation}
L=-\left((m_0+W)+W{\rm ln}\frac 1 {\sqrt{g_{\mu\nu}\dot x^\mu \dot
x^\nu}}\right)\sqrt{g_{\mu\nu}\dot x^\mu \dot x^\nu},
\label{2.27}\end{equation} where $\dot x^\mu=\frac d {d t} x^\mu$.
So the nonlinear term leads to a tiny departure from the geodesic.

For a spinor with interaction like electromagnetic field, the
mass-energy relation of the particle will be much complicated.
More generally, we consider the system with following
Lagrangian\cite{gu3}
\begin{equation}\begin{array}{ll} {\cal L}=&\psi^+\al^\mu
(i\pa_\mu-e A_{\mu})\psi-\mu\ck \ga+F(\ck\ga)-s\ck\ga G
\\& -\frac 1 2 \kp(\pa_\mu A_\nu\pa^\mu A^\nu-a^2 A_\mu A^\mu)-\frac 1 2
\la (\pa_\mu G\pa^\mu G-b^2 G^2),\end{array} \label{2.2.28}
\end{equation} where $A^\mu$ and $G$ are potentials produced by $\psi$ itself,
$\kp=\pm 1$ and $\la=\pm 1$ stand for the repulsive or attractive
self interaction. Then the 4-dimensional momentum $p^\mu$ and
energy $E$ of the system at particle state are respectively
defined by
\begin{eqnarray}
p^\mu&\equiv&\int_{R^3}\psi^{+}_{k}(i\pa^\mu-eA^\mu)\psi d^3x,\label{2.2.7} \\
E~ &\equiv& \int_{R^3}\left(\sum_{\forall f} \frac{\pa {\cal
L}}{\pa(\pa_t f)}\pa_t f-{\cal L}\right)d^3 x = p^0+E_F+E_A+E_G,
 \label{2.2.8}
\end{eqnarray}
where the classical parameters are given by
\begin{eqnarray}
p^{\mu} &=& \left(m_0+sW_\ga G+W_F{\ln}\frac 1 {\sqrt{1-v^2
}}\right)u^{\mu}, \label{2.2.25}\\
E_F&=&W_F\sqrt{1-v^2},\label{2.2.15}\\
E_A & =& W_A \frac 1{\sqrt{1-v ^2}}\left(1+\frac {v^2} 3 \right)+W_a\frac {v^2}{\sqrt{1-v^2}},\label{2.2.16} \\
 E_G &=& W_G \frac 1{\sqrt{1-v ^2}}\left(1-\frac {v^2} 3 \right)-W_b\frac
{v^2}{\sqrt{1-v^2}}, \label{2.2.17}
\end{eqnarray}
in which the proper parameters are calculated by
\begin{eqnarray}
W_F&=&\int_{R^3}(F'\ck\ga-F)d^3 \bar x>0,\quad W_\ga ~=~ \int_{R^3} \ck\ga d^3\bar x, \label{2.2.12}\\
W_A &=&\frac {\kp e^2}{8\pi}\int_{R^6}\frac {e^{-ar}} r\left(
{|\psi(\bar x)|^2| \psi(\bar y)|^2}+\vec\rho (\bar
x)\cdot\vec\rho(\bar
y)\right)d^3 \bar x d^3\bar y,\label{2.2.13}\\
W_G &=&-\frac {\la s^2}{8\pi}\int_{R^6}\frac {e^{-br}} r
{\ck\ga(\bar x)\ck\ga(\bar y)}d^3\bar x d^3\bar y, \label{2.2.14}\\
W_a&=&\frac \kp 3 \left( \frac {ae}{4\pi} \right)^2
\int_{R^3}\left(\int_{R^3} \frac {e^{-ar}} r |\psi(\bar y)|^2d^3
\bar y\right)^2 d^3 \bar x,\\
W_b&=&\frac \la 3 \left( \frac {bs}{4\pi} \right)^2
\int_{R^3}\left(\int_{R^3} \frac {e^{-br}} r \ck\ga(\bar y)d^3
\bar y\right)^2 d^3 \bar x,
\end{eqnarray}
and $r=|\bar x-\bar y|$. By (\ref{2.2.8})-(\ref{2.2.17}) we get
the dispersion relation as follows
\begin{eqnarray}
(E-E_F-E_A-E_G)^2=\vec p^{~2}+\left(m_0+sW_\ga G+W_F{\ln}\frac 1
{\sqrt{1-v^2 }}\right)^2. \label{2.2.24}
\end{eqnarray}
By (\ref{2.24}) and (\ref{2.2.24}), we find that the interaction
term can result in very complicated dispersion relation. How to
use these relation to explain the threshold paradoxes of the high
energy comic rays is involved in the interpretation of parameters,
which will be discussed elsewhere. By the way, the scalar
interaction may be absent in the nature, because it manifestly
appears in proper mass of feimions, see (\ref{2.2.25}) and
(\ref{2.2.24}).

\section{Discussion and conclusion}
\setcounter{equation}{0}

From the above analysis, we find that nonlinear field theories do
include the non-quadratic dispersion relation such as (\ref{2.24})
and (\ref{2.2.24}). So the explanation for threshold paradoxes of
UHECR and TeV $\ga$-rays by dispersion relation does not
definitely require Lorentz violation. It should be mentioned that,
all interactions are actually related to nonlinearity, for
instance, the charge density of the electromagnetic interaction
$\psi^+\al^\mu\psi$ is nonlinear, which contributes proper energy
$W_A$ for the particle as described by (\ref{2.2.13}). This part
of energy satisfy the energy-speed relation (\ref{2.2.16}).

By (\ref{2.2.25})-(\ref{2.2.17}), we learn different interaction
term leads to different energy-speed relation. So the experiments
towards such relations should bring us important information from
each interaction, and the high energy cosmic rays may be the
useful materials. On the other hand, the disturbance of nonlinear
effect may influence our astronomical observation. For the
movement of a heaven body, by (\ref{2.27}), we learn that the
order of the relative deviation from geodesic is $Err=\frac
{Wv^2}{mc^2}$, where $W$ is the energy contribution of all
interactions, $m$ is the usual mass. The typical values for a
proton are $W\sim 1$MeV and $m_p\sim 10^3$MeV, so we have $\frac W
m\sim 10^{-3}$. In a galaxy, the typical speed of heaven bodies
relative to the CMB is $300$km/s\cite{spd}, so we have the typical
nonlinear deviation for galactic system $Err|_{galaxy}\sim
10^{-9}$. In the solar system, the typical speed of the planets is
about $v\sim 30$km/s, so we have the typical nonlinear deviation
for solar system $Err|_{planet}\sim 10^{-11}$. Thus, before the
nonlinear effects are clearly worked out, an astronomical
measurement with relative error less than $Err$ is difficult. The
precession of the perihelion of Mercury is 43 seconds of arc per
century $43/(100\times360\times 3600)\approx 3.3\times 10^{-7}$,
so the nonlinear effects have not influence on this result.

Lorentz invariance includes two fold meanings: One is the
covariance of the universal physical laws, which is a problem of
philosophy. The other is property of the spacetime, namely the
measurement of line element. Whether the spacetime manifold is
measurable is also a philosophical problem, but how to measure the
distances is a problem of geometry. The philosophical problem
involves the most fundamental and universal postulates which could
only be acceptable as faith, because we can neither test all
particles whether they satisfy the covariant equation everywhere
and every time, nor can we check whether all parts of the
spacetime are measurable, including the singularity inside a black
hole and each points at a line of Planck length. Even though the
Lorentz violation theories can not manifestly violate the
covariance, see the forms of (\ref{1.6})-(\ref{1.8}) as example.
One may argue the thermodynamics is not covariant, the answer is
that it is just a conditional theory but not a universal one, thus
to abuse its concepts and laws, such as entropy and the second
law, without restriction is inadequate and leads to confusion.

We once measured the spacetime with rule $ds_t=|dt|,
ds_s^2={dx^2+dy^2+dz^2}$ and solved infinite practical problems.
Today we measure the universe with $ds^2=g_{\mu\nu}dx^\mu d x^\nu$
then we achieved beauty and harmony. There are infinite rules to
measure length consistently, but only the spacetime with rule of
quadratic form has wonderful properties and potential. If
accepting the quadratic rule, the local Lorentz invariance
certainly holds for the spacetime.

There are also fields satisfy the covariance but violate the
Lorentz invariance. Their Lagrangian always includes term as
follows\cite{LV}
$$
{\cal L}_{LV}=(g^{\mu\nu}+\tau^{\mu\nu})\pa_\mu \pa_\nu
\phi+\psi^+(\al^\mu+a^\mu)\pa_\mu \psi+\cdots.
$$
For such fields, the propagating speed is not pure geometrical,
which depends on fields $\tau^{\mu\nu}$ and $a^{\mu}$. Similar to
the case of Navier-Stockes equation in fluid mechanics, the
nonlinearity is much worse than that of (\ref{2.2.28}).  Although
for adequately small value $\tau^{\mu\nu}$ and $a^{\mu}$, the
solutions to the dynamical equation will also be finite, but the
world including such fields will become a mess, because each atom
of the same element have different spectrum depending on
coordinates, crystal lattices are distorted, the solar system has
turbulence and chaos, and the double-helix of DNA has disordered
knots. So {\bf `the coefficients of the highest order derivatives
in the Lagrangian must be constants'} should be a fundamental
postulate. This postulate is related to the quaternion structure
of the world, only of such excellent structure the world becomes
so luxuriant but so harmonious. In percipience of such opinion,
the modification of general relativity with terms $\phi R$, $f(R)$
is also doubtful.

Some confusions in physics are caused by ambiguous concepts or
circular relations\cite{qnt1,qnt2}. For example, to the relation
between classical mechanics(CM) and quantum mechanics(QM), the
common answer must be that: ``transform the classical parameters,
Hamiltonian and the energy equation of CM into operators, we get
QM. Contrarily, the limit of QM as $\hbar\to 0$ provides CM.''

At first, by the answer, it seems that both CM and QM are
equivalent theories in logic. In fact they are different theory
suitable for different status of a system. The basic concepts such
as `coordinate', `momentum' have different meanings in each
mechanics, although they have close relations. The uncertainty
relation is the typical plausible laws making puzzles and
paradoxes\cite{qnt1,qnt2}. Clearly they should be unified in a
higher level theory\cite{gu1,gu2,sln3,gu6}. secondly, $\hbar$ is a
universal constant acting as a unit to measure other physical
parameters, we have not a concept $\hbar\to 0$? Whether a physical
system should be described by CM or QM is obviously not determined
by external conditions such as $\hbar\to 0$ and CM or QM, but
determined by the status of the system itself. This is the meaning
of the {\bf Definition 2}.

Spacetime and fields are different components of the world. They
paly different roles with different characteristics, and satisfy
completely different postulates and measurement rules. So it is
hard to understand the motivation of the quantum gravity. Why we
should modify a well defined and graceful theory, without any
definite violation of experiments, by an ambiguous and incomplete
theory? Why not modify the quantum field theory by general
relativity? The mission of a physical theory is to find out the
intrinsic truth and beauty and harmony of the nature. But at its
best, besides some ill defined concepts as foam, wormhole, Lorentz
violation, what can quantum gravity actually provide us?

In the spinor theory of general relativity context, there exists
the {\bf pseudo-violation} of Lorentz invariance\cite{gu9,gu10},
which is caused by the derivatives of the vierbein or local frame.
The vierbein is defined in the tangent spacetime of a fixed point
in spacetime manifold, and the Lorentz transformation is just an
algebraic operation in this fixed tangent spacetime. Whereas the
derivatives of the vierbein must involve the tangent spacetimes of
different points in some sense, so it violates the local Lorentz
invariance. In strict sense, the equivalence principle only holds
for the linear tensors, where the influences of vierbein and
nonlinear fields are absent. However, the fundamental dynamic
equations of the field system should be intrinsically Lorentz
invariant.

\section*{Acknowledgments}

The author is grateful to his supervisor Prof. Ta-Tsien Li for his
encouragement and guidance.

\end{document}